\documentstyle[12pt]{article}
\begin{document} 
\begin{center}
{\Large On the Electromagnetic Response of Charged Bosons Coupled to a
Chern-Simons Gauge Field: A Path Integral Approach}
\vskip 7mm
{\large D.\ G.\ Barci \footnote{barci@symbcomp.uerj.br}} \\

{\it Universidade do Estado do Rio de Janeiro (UERJ)

Instituto de F\'\i sica, Departamento de F\'\i sica Te\'orica

R.\ S\~ao Francisco Xavier, 524 --- Maracan\~a --- CEP 20550-013

Rio de Janeiro, RJ --- Brazil.}
\vskip 5mm

{\large E.V.\ Corr\^ea Silva \footnote{ecorrea@symbcomp.uerj.br} }
 ~~~~~and~~~~~ {\large J.F.\ Medeiros Neto
 \footnote{medeiros@cat.cbpf.br}} \\

{\it Centro Brasileiro de Pesquisas F\'\i sicas (CBPF) 

Rua Xavier Sigaud 150 --- Urca --- CEP 22290--180 

Rio de Janeiro, RJ ---  Brazil. }
\vskip 5mm

April 30, 1998
\vskip 7mm

\end{center}

\begin{abstract}

We analyze the electromagnetic response of a system of charged bosons
coupled to a Chern-Simons gauge field. Path integral techniques are
used to obtain an effective action for the particle density of the
system dressed with quantum fluctuations of the CS gauge field. From
the action thus obtained we compute the $U(1)$ current of the theory
for an arbitrary electromagnetic external field. For the particular
case of a homogeneous external magnetic field, we show that the
quantization of the transverse conductivity is exact, even in the
presence of an arbitrary impurity distribution. The relevance of edge
states in this context is analyzed. The propagator of density
fluctuations is computed, and an effective action for the matter
density in the presence of a vortex excitation is suggested. 

\end{abstract} 

\newpage

\section{Introduction} 

Quantum Field Theory (QFT) techniques applied to Condensed Matter
systems have been extensively used in the recent past
\cite{livrofradking} \cite{kleinert}. Specially, path integral methods
have been applied successfully over the years in many problems of
Condensed Matter physics such as superconductivity, superfluidity,
plasmas, cristalization \cite{popov} and quantum Hall effect (QHE)
\cite{qhe}. The essence of this approach is the evaluation of an
effective partition function, obtained from the original one by
integration over some degrees of freedom. In general, this technique
leads to the problem of solving a fermion or boson functional
determinant, to be evaluated by some approximation scheme. An
application to the study of two-body interactions in the QHE can be
found in ref. \cite{bam}. Also, an interesting application to quasi
one-dimensional interfaces was developed in ref.\ \cite{bm2}. 

In the last years, models involving matter fields coupled to a $U(1)$
Chern-Simons gauge field were analyzed, not only as QFT models
\cite{c-s-c}, but also as fenomenological Condensed Matter
models\cite{c-s-cm}. The most interesting feature of these models is
that the Chern-Simons field attaches quantum fluxes to the particle
positions (or to the particle density distribution) thus changing the
{\em statistics} of the original fields. In particular, if we attach
an even number of quantum fluxes to a matter field, its statistics is
preserved (bosonic or fermionic). On the other hand, if we attach an
odd number of fluxes, the statistic of the relevant degrees of freedom
changes from bosonic to fermionic and vice-versa. This means that a
system of fermions can be regarded as {\em either} a system of bosons
with an odd number of attached fluxes {\em or} a system of (another
kind of) fermions plus an even number of fluxes attached to them. The
first possibility has been explored in the construction of the
Chern-Simons-Landau-Ginzburg (CSLG) theory for the Fractional Quantum
Hall Effect (FQHE)\cite{zang}; the second one has led to the
``composite fermion'' model for the FQHE \cite{jain}\cite{fradking}.

The electromagnetic response of the CSLG theory was calculated in ref.\
\cite{zang}. In that paper, the Hall conductivity is evaluated in the
saddle point approximation plus gaussian fluctuations. The essence of
this approximation scheme can be described as follows: first, the path
integral of charged bosons in the presence of an external gauge field is
evaluated in the gaussian approximation, yielding a quadratic effective
action for the external gauge field; then, the Chern-Simons coupling is
turned on, and the integral over the statistical gauge field is
computed. In this way, an interesting relation between bosonic
superconductivity and incompressibility of fermionic degrees of freedom
is established. 

Several authors argue that the quantization of the transverse
conductivity should be exact due to the topological character of the
Chern-Simons term. Nevertheless, to the best of our knowledge, a
complete explicit calculation, involving  random impurities, is
still lacking. In the case of the Integer Quantum Hall Effect (IQHE)
(where electron-electron interaction is not considered), it was shown
that the Hall conductivity is not affected by a delta-type impurity
\cite{prange}\cite{ricardo}. However, the study of a general type of impurities
combined with a Chern-Simons gauge coupling is a very difficult task
and no exact results are available.

Another aspect of these models is the study of density fluctuations
around the ground state. At the classical level, topological
\cite{boyan}\cite{zang2} \cite{ezawa} and non-topological
\cite{non-top} solutions to the equations of motion were found.
However, it would be interesting to study if the quantum fluctuations
of the gauge field do really modify the density profiles.

Motivated by these ideas, we will consider in this paper the
electromagnetic response of non-relativistic bosons coupled to a
Chern-Simons Gauge Field. To this aim, the path integral technique is
employed, so that all degrees of freedom of the model but  the
particle density are integrated out; hence, the resulting partition
function --- a functional integral over the charge density alone ---
takes into account the quantum fluctuations of the gauge field. (The
charge density is, perhaps, the simplest of gauge invariant objects.)
In particular, the Chern-Simons constraint (that attaches fluxes to
particles) is implemented exactly, rather than being artificially
imposed by means of some approximation scheme.

This technique allows us to clarify some aspects of this system.
First, we can write an explicit expression for the $U(1)$ current of
the model submitted to an {\em arbitrary} electromagnetic field.
Specializing this general expression for the current to the case of a
static and homogeneous external magnetic field, we can explicitly show
that the quantization of the transverse conductivity is exact, {\em
even in the presence of any type of impurities}. Thus, we generalize
Prange's result \cite{prange}\cite{ricardo} to the FQHE and to any 
kind of impurities in the context of the CSLG theory.

This paper is organized as follows: In $\S\ref{CSB}$ we deduce the
effective action for the bosons coupled to a Chern-Simons field plus
an arbitrary electromagnetic field, in the density representation. In
$\S\ref{ER}$ we evaluate the $U(1)$ current of the model. We show the
exact quantization of the Hall conductivity and the role played by the
edge states. In $\S\ref{DF}$ we analyze the dynamics of density
fluctuations around the ground state. Finally we discuss our results
and present our conclusions in $\S\ref{dis}$.

\section{The Effective Action for Bosons Coupled to a Chern-Simons
Gauge Field in the Density Representation. 
\label{CSB} }

Let us begin by considering the Euclidean action for non-relativistic
bosons coupled to a $U(1)$ Chern-Simons gauge field $a_\mu(x)$ and an
{\em arbitrary} external electromagnetic field $A_\mu(x)$, 

\begin{eqnarray}
S&=&\int d^2xd\tau~\left\{  \phi^*\left( \partial_\tau+i[A_0+a_0]-
i\mu\right)\phi-\frac{1}{m}\left|\left( \frac{1}{i}\vec{\nabla}-e 
(\vec{A}+\vec{a})\right)\phi \right|^2 \right\} +
\nonumber\\
&-&\frac{1}{2}\int d^2xd^2y d\tau~ \left(\phi^*(x)\phi(x)-
\bar{\rho}\right)V(x-y)
\left(\phi^*(y)\phi(y)-\bar{\rho}\right) +
\nonumber \\
&-&\frac{i}{2}\left(\frac{\pi}{\theta}\right)\int d^2x d\tau ~
\epsilon^{\mu\nu\rho}a_\mu\partial_\nu a_\rho~~~~~~~~~~~~~~~~. 
\label{1}
\end{eqnarray}
where $V(x-y)$ is an arbitrary two-body potential, and the last term
is the well-known Chern-Simons term. Due to this term, the classical
equation of motion related to $a_0$ 

\begin{equation}
\rho(x)=\phi^*\phi=\frac{\pi}{\theta}\vec{\nabla}\times a~~~ 
\end{equation}
is a pure constraint that attaches $\frac{\theta}{\pi}$ fluxes to the
particle density. In order to represent fermionic degrees of freedom
from the bare bosonic fields, we must choose $\theta=(2n+1)\pi$, with
integer $n$. 

The partition function is 

\begin{equation}
Z(A_0,A_i)=
\int {\cal D}\phi^*{\cal D}\phi{\cal D}a_\mu ~ G_F(\phi,a_{\mu})~~ 
e^{-S(\phi,\phi^*,a_\mu, A_\mu)}
\label{2}
\end{equation}
where $G_F(\phi,a_{\mu})$ is a gauge fixing functional, necessary to
avoid overcounting of physical states in the functional integral.

The presence of the Chern-Simons field allows us to extract the external
electromagnetic field $A_\mu$ from the kinetic operator of the action. 
Making the following  change of variables:
\begin{eqnarray}
a_\mu &\longrightarrow& a_\mu-A_\mu  \nonumber \\
{\cal D}a_\mu &\longrightarrow&   {\cal D}a_\mu 
\label{3}
\end{eqnarray}
we obtain
\begin{eqnarray} 
S&=&\int d^2xd\tau~\left\{ \phi^* \left(
\partial_\tau+ia_0-i\mu\right)\phi- \frac{1}{m}\left|\left(
\frac{1}{i}\vec{\nabla} +\vec{a}\right)\phi \right|^2 \right\}
\nonumber\\ 
&-&\frac{	1}{2}\int d^2x d^2y d\tau~
\left(\phi^*(x)\phi(x)-\bar{\rho}\right)V(x-y)
\left(\phi^*(y)\phi(y)-\bar{\rho}\right) \nonumber \\
&-&\frac{i}{2}\frac{\pi}{\theta}\int d^2x d\tau ~
\epsilon^{i0j}a_i\partial_0 a_j -i\frac{\pi}{\theta}\int d^2x d\tau ~a_0
(\vec{\nabla}\times\vec{a}+B) \nonumber \\ &-&i \frac{\pi}{\theta}\int
d^2x d\tau ~ \epsilon^{ij}a_i E_j -\frac{i}{2}\frac{\pi}{\theta}\int
d^2x d\tau ~ \epsilon^{\mu\nu\rho}A_\mu\partial_\nu A_\rho \label{4}
\end{eqnarray} 
where the matter field couples to the dynamical gauge field only.
Here we use $\vec{E}$ and $B$ to denote the external electric and
magnetic field, respectively. (In what follows, greek indices run from
$0$ to $2$ and latin indices run from 1 to 2.)

Going over the density representation, it is useful to perform another
change of variables,
\begin{eqnarray}
\phi(x)&=&\rho^{1/2}(x)~ e^{i\theta(x)} \nonumber \\
\phi^*(x)&=&\rho^{1/2}(x)~ e^{-i\theta(x)}
\label{rhotheta}
\end{eqnarray}
with  trivial Jacobian
\begin{equation}
{\cal D}\phi{\cal D}\phi*={\cal D}\rho {\cal D}\theta
\end{equation}
Replacing (\ref{rhotheta}) in (\ref{4}) we have
\begin{eqnarray}
S&=&\int d^2xd\tau~\left\{
\frac{1}{2}\partial_\tau\rho+
i\rho(\partial_\tau\theta+a_0)\right\}
-\frac{1}{2 m}\left\{ \frac{1}{4\rho}
\vec{\nabla}\rho\cdot\vec{\nabla}\rho+\rho 
\left|\vec{\nabla}\theta-\vec{a}\right|^2  \right\}
 \nonumber\\
&-&\frac{1}{2}\int d^2x d^2y d\tau~ 
\left(\rho(x)-\bar{\rho}\right)V(x-y)
\left(\rho(y)-\bar{\rho}\right)+\mu\int d^2x d\tau \rho(x) 
\nonumber \\
&-&\frac{i}{2}\frac{\pi}{\theta}\int d^2x d\tau ~
\epsilon^{i0j}a_i\partial_0 a_j-i\frac{\pi}{\theta}
\int d^2x d\tau ~a_0
(\vec{\nabla}\times\vec{a}+B) 
\nonumber \\
&-&i \frac{\pi}{\theta}\int d^2x d\tau ~
\epsilon^{ij}a_i E_j -\frac{1}{2}\frac{\pi}{\theta}
\int d^2x d\tau ~
\epsilon^{\mu\nu\rho}A_\mu\partial_\nu A_\rho ~~~~~~~~~~~. 
\label{6}
\end{eqnarray}

The idea  is to integrate out the gauge fields $a_0(x),a_i(x)$ and
the phase $\theta(x)$, in order to obtain an effective action in
terms of the density $\rho(x)$ only. At this point, it is useful to
discuss the gauge fixing functional $G_F(\phi,a_\mu)$.

There is great freedom in choosing $G_F(\phi,a_\mu)$, and the
partition function $Z(A_0,A_i)$ must be gauge invariant, i.e., it must
not depend on any particular choice. Notice that in the ``polar
coordinates'' we are using, a gauge transformation reads

\begin{eqnarray} 
\theta(x)&\longrightarrow & \theta(x)+\Lambda(x) \label{gauge1}\\
a_\mu(x)&\longrightarrow & a_\mu(x)+\partial_\mu\Lambda(x)~~~~~~~~.
\label{gauge2} 
\end{eqnarray} 
So, we can always choose $\Lambda(x)=-\theta(x)$ to have $\theta(x)=0$
in the action. This corresponds to choosing the fixing functional to
be
\begin{equation}
G_F(\theta,a_\mu)=\delta\left(\theta(x)\right)~~~~~~~~~.
\label{GFT}
\end{equation}
We must prove that this is an accessible gauge, and that it does
completely fix the gauge. It is not difficult to show \cite{kleinert}
that this gauge is equivalent to Coulomb's gauge 
\begin{equation}
G_F(\theta,a_\mu)=\delta\left(\vec{\nabla}\cdot \vec{a}\right).
\label{cg}
\end{equation}
Demonstrating such equivalence consists of showing the existence of a
continuous gauge transformation that leads from condition (\ref{GFT})
to (\ref{cg}). Decomposing the field $a_i$ into a transversal $a_i^\perp$
plus a longitudinal component $a_i^L$, it is possible to rewrite the
integration measure in the Coulomb's gauge in the following form, 
\begin{equation}
\int {\cal D}a_i  {\cal D}\rho {\cal D}\theta~~ 
\delta\left(\vec{\nabla}\cdot \vec{a} \right)\ldots =
\int {\cal D}a^{\perp}  {\cal D}\rho {\cal D}\theta\ldots  
\label{tl}
\end{equation}
as the Coulomb's gauge sets the longitudinal component of the 
gauge field to zero. 
Let us now perform a gauge transformation
\begin{equation}
a_i(x)\longrightarrow  a_i(x)-\partial_i\theta(x)~~.  
\end{equation}
This transformation removes the phase of $\phi(x)$ from the action
but adds a longitudinal component $a^L_i(x)=-\partial_i\theta(x)$ to 
$a^{\perp}_i(x)$.
Then, we can replace ${\cal D}\theta$ with ${\cal D}a^L(x)$ in
(\ref{tl}) (with a trivial Jacobian) obtaining

\begin{equation}
(\ref{tl})=\int {\cal D}a^{\perp} {\cal D}a^L(x) {\cal D}\rho \ldots=
\int {\cal D}a_i  {\cal D}\rho {\cal D}\theta~~   
\delta\left(\theta(x) \right)\ldots 
\label{fixing}
\end{equation}
In this way, we show that (\ref{GFT}) and (\ref{cg}) are equivalent.

Now, we are ready to begin with the integration. We see that equation
(\ref{6}) has only linear terms in $a_0(x)$. Integration is trivially
carried out, yielding the well-known Chern-Simons constraint
\begin{equation}
\int{\cal D}a_0~~ e^{-i\int d^2x d\tau~a_0(\rho-\frac{\pi}{\theta}
[\vec{\nabla}\times\vec{a}+B]) }=\delta(\rho-\frac{\pi}{\theta}
[\vec{\nabla}\times\vec{a}+B])~~~~~~~~.
\label{7}
\end{equation}
The quantization of this model in the framework of constrained systems
has been considered by several authors; in particular, we find the
gauge invariant approach of reference \cite{boyan} very instructive.
However, in the present paper, we prefer to follow a functional
approach, i.e., we will integrate over all configurations of $\rho$
and $a_i$ that satisfy
\begin{equation}
\rho(x)=\frac{\pi}{\theta}
[\vec{\nabla}\times\vec{a}+B]. 
\label{csc}
\end{equation} 
Note that the attachment of magnetic fluxes to the particle density
$\rho(x)$ is {\em not} a classical effect. It remains in all the
integration over $\rho$ and $a_i$. A very interesting physical
realization of this phenomenon can be found in \cite{vortices}. Thus,
using (\ref{7}) and (\ref{GFT}) we find for the partition function
\begin{equation}
Z(A_0,A_i)=\int {\cal D}\rho{\cal D}\vec{a}~\delta
\left(\rho-\frac{\pi}{\theta}
[\vec{\nabla}\times\vec{a}+B]\right) ~~e^{-S_F}
\label{8}
\end{equation}
with 

\begin{eqnarray}
S_F&=&-\frac{1}{2 m}\int d^2xd\tau~
\left\{ \frac{1}{4\rho}\vec{\nabla}\rho\cdot\vec{\nabla}\rho+
\rho|\vec{a}|^2  \right\} +\mu\int d^2x d\tau \rho(x)
\nonumber\\
&-&\frac{1}{2}\int d^2x d^2y d\tau~ 
\left(\rho(x)-\bar{\rho}\right)V(x-y)
\left(\rho(y)-\bar{\rho}\right) 
\nonumber \\
&-&\frac{i}{2}\frac{\pi}{\theta}\int d^2x d\tau ~
\epsilon^{i0j}a_i\partial_0 a_j
-i \frac{\pi}{\theta}\int d^2x d\tau ~
\epsilon^{ij}a_i E_j 
\nonumber \\
&-&\frac{i}{2}\frac{\pi}{\theta}\int d^2x d\tau ~
\epsilon^{\mu\nu\rho}A_\mu\partial_\nu A_\rho 
\label{9}
\end{eqnarray}
Clearly, $S_F$ is not gauge invariant, as we have fixed the gauge to
perform the integrals. Of course, {\em after the integration}, this
symmetry must be restored. Another observation about $S_F$ is that the
density $\rho(x)$ has no independent dynamics; rather, all of its
dynamics depends on its coupling to the gauge field. This is an
important fact in the study of the dynamic of density fluctuations. We
will return to this point in section \S\ref{DF}. 

The next step of our development is to integrate out the field $a_i$,
to obtain an effective action only for $\rho(x)$. In order to do that,
it is useful to decompose the bidimensional vector $a_i$ into a
longitudinal plus a transversal part

\begin{equation}
a_i=\partial_i\varphi+\epsilon_{ij}\partial_j\eta~~~~~~~~~~~~~. 
\label{decomp}
\end{equation}
This is a linear transformation. So, 
\begin{equation}
{\cal D}a_1{\cal D}a_2\longrightarrow{\cal D\eta}{\cal D\varphi}
\end{equation}
This decomposition is equivalent to describing a vector field
through its rotational and divergence, since

\begin{eqnarray}
\vec{\nabla}\cdot\vec{a}&=&\nabla^2\varphi     \label{div} \\
\vec{\nabla}\times\vec{a}&=&-\nabla^2\eta       \label{rot}       
\end{eqnarray}
Replacing (\ref{decomp}) in (\ref{9}) we find
\begin{eqnarray}
S_F&=&-\frac{1}{2 m}\int d^2xd\tau~
\left\{ \frac{1}{4\rho}\vec{\nabla}\rho\cdot\vec{\nabla}\rho\right\} 
+\mu\int d^2x d\tau \rho(x)
\nonumber\\
&-&\frac{1}{2}\int d^2x d^2y d\tau~ \left(\rho(x)-\bar{\rho}\right)V(x-y)
\left(\rho(y)-\bar{\rho}\right) 
\nonumber \\
&-&\frac{1}{2m}\int d^2x d\tau~ \rho(x)
\left(\vec{\nabla}\varphi\cdot\vec{\nabla}\varphi
+\vec{\nabla}\eta\cdot\vec{\nabla}\eta \right)
-i\frac{\pi}{\theta}\int d^2xd\tau~\varphi\partial_0\nabla^2\eta 
\nonumber \\
&+&i\frac{\pi}{\theta}\int d^2xd\tau~\varphi(\vec{\nabla}\times\vec{E})+
\eta(\vec{\nabla}\cdot\vec{E})
\nonumber \\
&-&\frac{i}{2}\frac{\pi}{\theta}\int d^2x d\tau ~
\epsilon^{\mu\nu\rho}A_\mu\partial_\nu A_\rho 
\label{17}
\end{eqnarray}
This action is quadratic in the field $\varphi$. Moreover, the
$\delta-\mbox{functional}$ in (\ref{8}) depends only on $\eta$.
Therefore, we can integrate over $\varphi$ using the relation
\begin{equation}
\int {\cal D}\varphi e^{-\int dx
~\frac{1}{2}\varphi(x)\hat{O}\varphi(x)+J(x)\varphi(x) }= 
Det^{-1/2}(\hat{O}) e^{\frac{1}{4}\int dx dy~J(x) \hat{O}^{-1}(x-y) J(y)}
~~~~, 
\end{equation}
thus obtaining 
\begin{equation}
Z(A_0,A_i)=\int {\cal D}\rho{\cal D}\eta~\delta(\rho-\frac{\pi}{\theta}
[B-\nabla^2\eta])~~e^{-S'_F}
\label{part}
\end{equation}
with
\begin{eqnarray}
S'_F&=&-\frac{1}{2 m}\int d^2xd\tau~
\left\{ \frac{1}{4\rho}\vec{\nabla}\rho\cdot\vec{\nabla}\rho\right\} 
+\mu\int d^2x d\tau \rho(x)
-\frac{1}{2}Tr\ln(\vec{\nabla}\cdot\frac{\rho}{m}\vec{\nabla})
\nonumber\\
&-&\frac{1}{2}\int d^2x d^2y d\tau~ \left(\rho(x)-\bar{\rho}\right)V(x-y)
\left(\rho(y)-\bar{\rho}\right) \nonumber \\
&-&\frac{1}{2m}\int d^2x d\tau~ \rho(x)
\vec{\nabla}\eta\cdot\vec{\nabla}\eta 
\nonumber \\ 
&+&\frac{m}{4}\frac{\pi^2}{\theta^2}\int d^2x d^2yd\tau~
(\partial_0\nabla^2\eta(x)-\vec{\nabla}\times\vec{E}(x))
\frac{1}{\vec{\nabla}\cdot(\rho\vec{\nabla})}
(\partial_0\nabla^2\eta(y)-\vec{\nabla}\times\vec{E}(y)	) 
\nonumber \\
&+&i\frac{\pi}{\theta}\int d^2xd\tau~\eta(\vec{\nabla}\cdot\vec{E})
-\frac{i}{2}\frac{\pi}{\theta}\int d^2x d\tau ~
\epsilon^{\mu\nu\rho}A_\mu\partial_\nu A_\rho 
\label{19}
\end{eqnarray}
where we have used the fact that $\ln Det\hat O=Tr\ln\hat O$. 

Finally, we must integrate over the transversal component of $a_i$,
say $\eta(x)$. This is not a difficult task because $\eta(x)$ is in
the argument of a $\delta-\mbox{functional}$ (see eq. (\ref{7})). From
(\ref{csc}) we see that the support of the $\delta- \mbox{functional}$
consists in all the field configurations that satisfy
\begin{equation}
\nabla^2\eta=B-\frac{\theta}{\pi}\rho
\end{equation}
We can invert this linear differential equation by using 
 
\begin{equation}
\eta(x)=-\frac{\theta}{\pi}\int d^2y~G(x-y)(\rho(y)-\frac{\pi}{\theta}B(y))
\label{11}
\end{equation}
where $G(x-y)$ is the Green function for the Laplacian operator,
\begin{equation}
\nabla^2G(x-y)=\delta(x-y)
\label{12}
\end{equation}
The well-known property of the $\delta-\mbox{function}$
\begin{equation}
\delta(f(x))=\sum \frac{\delta(x-x_i)}{|f'(x_i)|}, ~~~~f(x_i)=0. 
\end{equation}
can be extended to the  $\delta-\mbox{functional}$, yielding
\begin{equation}
\delta\left(\rho-\frac{\pi}{\theta}(B-\nabla^2\eta)\right)
=\frac{1}{|\nabla^2\delta|}
\delta\left(\eta(x)+\frac{\theta}{\pi}\int d^2y~G(x-y)
\left\{\rho(y)-\frac{\pi}{\theta}B(y)\right\}
\right)
\label{13}
\end{equation}
The factor $|\nabla^2\delta|$ is field-independent and can be absorbed
in the global normalization constant of the partition function. Using
(\ref{13}), it is now simple to functionally integrate out the field
$\eta$, obtaining
\begin{equation}
Z(A_0,A_i)=e^{\frac{i}{2}\frac{\pi}{\theta}\int d^2x d\tau ~
\epsilon^{\mu\nu\rho}A_\mu\partial_\nu A_\rho }
\int {\cal D}\rho~~ e^{-S_{eff}(\rho,\vec{E},B)}
\label{22a}
\end{equation}
with
\begin{eqnarray}
S_{eff}&=&
\frac{m}{2}\int d^3x d^3y~
\partial_0\rho\left\{
\frac{1}{\vec{\nabla}\cdot(\rho\vec{\nabla})}  \right\}
\partial_0\rho
-\frac{1}{2 m}\int d^2xd\tau~
\left\{ \frac{1}{4\rho}
\vec{\nabla}\rho\cdot\vec{\nabla}\rho \right\} \nonumber \\
&-&\frac{1}{2}\int d^2x d^2y d\tau~ 
\left(\rho(x)-\bar{\rho}\right)V(x-y)
\left(\rho(y)-\bar{\rho}\right)
-\frac{1}{2}Tr\ln(\vec{\nabla}\cdot\frac{\rho}{m}\vec{\nabla})
\nonumber \\
&-&\frac{1}{2}\int d^2x d^2y d\tau~ 
\left(\rho(x)-\frac{\pi}{\theta}B(x)\right)F(x-y)
\left(\rho(y)-\frac{\pi}{\theta}B(y)\right) 
\nonumber \\
&-&\frac{1}{2m}\frac{\theta^2}{\pi^2}\int d^2xd^2yd^2zd\tau~
\vec{\nabla}G(x-y)\cdot\vec{\nabla}G(x-z)\times \nonumber \\
&&~~~~~~~~\times\left(\rho(x)-\frac{\pi}{\theta}B(x)\right)
\left(\rho(y)-\frac{\pi}{\theta}B(y)\right)\left(\rho(z)-
\frac{\pi}{\theta}B(z)\right)
\nonumber \\
&-&\int d^2xd^2yd\tau~\left(\rho(x)-\frac{\pi}{\theta}B(x) \right)
G(x-y)\vec{\nabla}\cdot\vec{E}(y)   
\label{22}
\end{eqnarray}
where
\begin{equation}
F(x-y)=\frac{\theta}{\pi}\frac{1}{m}
\int d^2z~ B(z)\vec{\nabla}G(z-x)\cdot
\vec{\nabla}G(z-y)
\end{equation}

Equations (\ref{22a}) and (\ref{22}) are the main results of this
section. $S_{eff}$ is the effective action for the Chern-Simons Bosons
in the density representation, interacting with an {\em arbitrary
electromagnetic field}. This is a gauge invariant action, as it only
depends on $\rho(x)$, $\vec{E}$ and $B$. The coupling of the charge
density to the electromagnetic field is very peculiar. The magnetic
field acts as a ``background density'', entering the action only
through terms of the form
$\left(\rho(x)-\frac{\pi}{\theta}B(x)\right)$. The electric field
couples to the density only through its divergence
$\vec{\nabla}\cdot\vec{E}$, which is obviously proportional to the
{\em external charges} (impurities, for example). So the last term of
(\ref{22}) is simply the two-dimensional Coulomb energy between the
external charges and the charge fluctuations over the magnetic
background. In this term we absorbed the chemical potential as it can
be simulated by a uniform background charge density.  
Note that, in particular, a homogeneous electric field do not couple to
the matter field. 

The coupling of the gauge field with the bosonic matter field had two
main consequences in the particle density of the system. The
integration over the {\em longitudinal} gauge field produced a
non-local dynamical term for $\rho(x)$ (see first term of (\ref{22}))
and a non-local interaction term given by
$Tr\ln(\vec{\nabla}\cdot\frac{1}{m}\rho\vec{\nabla})$. The integration
over the {\em transversal} gauge field induced a two-body interaction
(third term of (\ref{22})) and a three-body interaction also. The
induced two-body interaction is the basic characteristic of the
fluctuation of the transversal degrees of freedom of Chern-Simons
field and, for the particular case of
$B(z)\equiv B=const.$, 
\begin{equation}
F(x-y)=-\frac{\theta}{\pi}\frac{B}{m}G(x-y)~~~~~~~, 
\end{equation}
we obtain the well-known logarithmic two-body interaction, extensively
explored in the past. This interaction is responsible for opening a
gap in the spectrum of excitations over the ground state of the
system. Moreover, the longitudinal fluctuations are non-trivial. In
particular, the non-local dynamical term will modify the propagator of
the density fluctuations. Also, we note that a Chern-Simons term for
the {\em external electromagnetic field} has been factored out from the
action. In the next section we will take advantage of these properties
to analyze the electromagnetic response of this system. 

\section{Electromagnetic Response        \label{ER}}

This section is devoted to the calculation of the electromagnetic
response of the theory described in the last section.

Due to  gauge symmetry,  the  conserved current is  given by

\begin{equation}
<J_\mu(x)>=\frac{\delta~}{\delta A_\mu(x)}\ln(Z)~~~~.
\label{current}
\end{equation}
Explicitly evaluating the functional derivatives, we find using
(\ref{22a}) and (\ref{22}):
\begin{eqnarray}
<J_0(x)>&=& i\frac{\pi}{\theta}B+
\partial_i<\frac{\delta S_{eff}(\vec{E},B)}{\delta E_i})> 
\label{36} \\
<J_i(x)>&=& -i\frac{\pi}{\theta} \epsilon_{ij} E_j  +
\partial_0<\frac{\delta S_{eff}(\vec{E},B)}{\delta E_i})>
+\epsilon_{ij}\partial_j <\frac{\delta S_{eff}(\vec{E},B)}{\delta B})>
\label{35} 
\end{eqnarray}
As $S_{eff}$ is gauge invariant, it depends on $A_i$ only through
$\vec{E}$ and $B$. This implies the existence of a topological
current, automatically conserved (last term of (\ref{35})). We will
see that this topological current is responsible for the exact
quantization of the Hall conductivity, even in the presence of
impurities, a fact closely related to the gauge invariance and edge
state excitations. 

We can easily evaluate expression (\ref{36}), obtaining the obvious
result

\begin{equation}
<iJ_0>=<\rho(x,\tau)>
\label{44}
\end{equation}
The calculation of (\ref{35}) is  direct, but the result is less
trivial:

\begin{eqnarray}
<J_i>&=&-i\frac{\pi}{\theta} \epsilon_{ij}
\left( E_j-\partial_j\int d^2y~G(x-y)
\vec{\nabla}\cdot\vec{E}\right) \nonumber \\
&+&i\int d^2y~ \partial_i G(x-y)\partial_0
<\rho(y)-\frac{\pi}{\theta}B(y)>\nonumber \\
&+& J^T_i(x)
\label{Jexacta}
\end{eqnarray}
where

\begin{equation}
J^T_i(x)=\frac{1}{m}\frac{\theta}{\pi}\epsilon_{ij}
\partial_j \Delta(x)
\label{top}
\end{equation}
and

\begin{equation}
\Delta(x)=
\int d^2zd^2y~  \vec{\nabla}G(x-y)\cdot\vec{\nabla}_yG(y-z)
<(\rho(y)-\frac{\pi}{\theta}B(y))(\rho(z)-\frac{\pi}{\theta}B(z))>
\label{delta}
\end{equation}

This is the first result of this section. The current (\ref{Jexacta})
is the response of the system to an arbitrary electromagnetic field.
Before we go over to the calculation, it is necessary at this point to
check the conservation of the current (\ref{Jexacta}). So, let us
evaluate the divergence of $<J_i>$. The last and the second terms of
the first line of (\ref{Jexacta}) are automatically conserved because
both of them are {\em topological terms}. That means, they are of the
form $\epsilon_{ij}\partial_j f(x)$, with arbitrary $f$. The other
part of the current needs some attention. From (\ref{Jexacta}) we have
\begin{equation}
\partial_i<J_i>=-i\frac{\pi}{\theta} \epsilon_{ij}\partial_iE_j 
+i\int d^2y~ \nabla^2 G(x-y)\partial_0<\rho(y)-\frac{\pi}{\theta}B(y)>
\label{div1}
\end{equation}
Using (\ref{12}) we can integrate the last term obtaining
\begin{equation} 
\partial_i<J_i>= i\partial_0<\rho(y)>-i\frac{\pi}{\theta}
\left(\vec{\nabla}\times \vec{E} +\partial_0 B\right) \label{div2} 
\end{equation} 
From (\ref{44}) (remembering that the external electromagnetic field
satisfies Faraday's Law), we obtain the current conservation law (in
Euclidean space)
\begin{equation}
\partial_i<J_i>+\partial_0 <J_0>=0
\label{conservation}
\end{equation}

For an arbitrary electromagnetic field, 
expression (\ref{Jexacta}) can turn out to be quite a
complicated one.
However, it is useful for the study of the structure of the current for
specific configurations of the electromagnetic field. 
For example,  let us analyze the case of a static and uniform 
magnetic field, and a static but {\em arbitrary electric field}.
This case is the relevant one in the study of
the Quantum Hall Effect. 
In this case, equation (\ref{Jexacta}) reduces to 
\begin{eqnarray}
<J_i>&=&-i\frac{\pi}{\theta} \epsilon_{ij}
\left( E_j-\partial_j\int d^2y~G(x-y)
\vec{\nabla}\cdot\vec{E}\right) \nonumber \\
&+& J^T_i(x)
\label{Jexactaunif}
\end{eqnarray}
The first line of (\ref{Jexactaunif}) does not depend on any specific
detail of the system. Clearly, the dynamical properties of the system
(two-body interactions, etc.) are all contained in the topological
current $J_i^T(x)$. So, let us analyze this current more carefully. From
(\ref{delta}), we have 
\begin{equation} 
\Delta(x)= \int d^2zd^2y~
\vec{\nabla}G(x-y)\cdot\vec{\nabla}_yG(y-z) <\delta\rho(y)\delta\rho(z)>
\label{delta1} 
\end{equation}
where the expectation value is evaluated with the partition function
(\ref{22a}), and $\delta\rho$ is the density fluctuation around the
magnetic background. The action $S_{eff}$ only depends on $\vec{E}$
through its divergence. So, in a clean sample (without external
charges), $\Delta(x)$ does not depend on the electric field. Moreover,
we have that $<\rho>$ is constant due to translation invariance and
$<\rho>=\bar{\rho}$ fixes charge neutrality. Also, we saw that the
effect of the quantum fluctuations of the Chern-Simons field was that
of inducing a new two-body and three-body interactions with
neutralizing background $\frac{\pi}{\theta}B$. In order to define the
density fluctuations $\delta\rho$ of equation (\ref{delta1})
correctly, we must have 
\begin{equation}
\bar{\rho}=\frac{\pi}{\theta}B 
\end{equation} 
thus implying a Landau filling factor $\nu=\frac{\pi}{\theta}$. In
this case the two-point correlation function of density fluctuations
$<\delta\rho(x)\delta\rho(y)>$ depends on $|x-y|$. In the case of a 
sample with impurities, we must calculate the average of the 
correlation function with some weight factor (gaussian, for example). 
Thus, upon averaging the random impurities we restore the translation 
invariance and
\begin{equation}
\overline{<\delta\rho(x)\delta\rho(y)>}=\sigma(x-y) 
\end{equation} 
(the bar means ``average over random impurities''). 
In any case, we can write for $\Delta$, 
\begin{eqnarray} 
\overline{\Delta(x)}&=& \int d^2zd^2y~
\vec{\nabla}G(x-y)\cdot\vec{\nabla}_yG(y-z) \sigma(y-z) \nonumber \\
&=&\int d^2y~
\vec{\nabla}G(x-y)\cdot\vec{\nabla}_y \int d^2z G(y-z) \sigma(y-z) \nonumber \\
\label{delta2} 
\end{eqnarray}
In an infinite plane, the last integral of (\ref{delta2}) is constant
and therefore  $\overline{\Delta(x)}=0$, for $\nu=\pi/\theta$. 
So, in an infinite system the topological current $j^T_i$ given by
(\ref{top}) is zero. 

In a finite plane, $j^T_i$ is no longer zero, but we will show that it
does not contribute to the total current. To see this, let us write
the total current due to the topological density current as
\begin{equation}
I^{T}_i=\int_D j^T_i~ dS=
\frac{1}{m}\frac{\theta}{\pi}\int_D\epsilon_{ij}\partial_j \Delta(x)
~ dS
\end{equation}
The domain of integration $D$ is simply the surface of the sample.
Supose that $D$ is limited by a border $\partial D$. In this case we can
represent the function $\Delta$ as $\Delta(x)\Theta(D)$, where
$\Theta(D)$ is one in $D$ and zero otherwise. So, 
\begin{eqnarray}
\partial_j\left(\Delta(x)\Theta(D)\right)&=&\partial_j\Delta(x)\Theta(D) +
\Delta(x)\partial_j\Theta(D) \nonumber \\
&=&\partial_j\Delta(x)\Theta(D)- \hat{n}_j \Delta(x) \delta(\partial D)
\end{eqnarray} 
where $\delta(\partial D)$ is the Dirac delta-functional with support
in $\partial D$, and $\hat{n}_j $ is the $j$-th component of the
versor contained in the plane, normal to $\partial D$ and external to
$D$. 
Hence,
\begin{eqnarray}
I^{T}_i&=&\frac{1}{m}\frac{\theta}{\pi}
\left\{ \int\epsilon_{ij}\partial_j \Delta(x)\Theta(D)~ dS-
\int\epsilon_{ij}\hat{n}_j\Delta(x)\delta(\partial D) \right\}
\nonumber \\
&=&\frac{1}{m}\frac{\theta}{\pi}
\left\{\oint_{\partial D} d\vec{l}~ \hat{t_i}\Delta(x) 
-\int\epsilon_{ij}\hat{n}_j\Delta(x)\delta(\partial D)~ds\right\} 
\end{eqnarray} 
where $t_i=\epsilon_{ij}\hat{n}_j$ is a tangent versor to $\partial D$. 
But, 
\begin{equation}
\int\epsilon_{ij}\hat{n}_j\Delta(x)\delta(\partial D)~ds=
\oint_{\partial D} d\vec{l}~ \hat{t_i}\Delta(x) 
\label{55}
\end{equation}
which implies that $I^{T}_i=0$.  

The r.h.s of (\ref{55}) represents an edge current, responsible for
the cancellation of the topological total current. It means that, due
to the edge currents, the main contribution to the total current comes
only from the first line of (\ref{Jexacta}).
The relevance of edge states in the QHE was first stressed in 
\cite{halpering2} and then, its theory was developed in refs.\ 
\cite{stone}\cite{wen1}\cite{wen2}.  

It is interesting to note that, for $\nu\neq \pi/\theta$, there are two
different ``uniform backgrounds'', say, $\bar\rho$ and $(\pi/\theta) B$. In
this case, it is not possible for the mean value of the density being
constant. The translation invariance is broken and the density
fluctuations  no longer depend simply on $|x-y|$, but show a non-trivial
dependence on $x$ and $y$. This shows that for $\nu\neq \pi/\theta$, the
current will not  be given by the first line of (\ref{Jexacta}),
but the topological density current $J_i^T$ will contribute in a
non-trivial way. 

Let us analyze now the main part of (\ref{Jexacta}), namely
\begin{equation}
<J_i>=-i\frac{\pi}{\theta} \epsilon_{ij}
\left( E_j-\partial_j\int d^2y~G(x-y)
\vec{\nabla}\cdot\vec{E}\right) 
\label{Jhall1}
\end{equation}
Its second term has support only in the region $\rho(x)\neq 0$ and
depends on the divergence of $\vec{E}$, only. Thus, in a clean sample,
$\vec{\nabla}\cdot\vec{E}=0$ and the current is the well-known Hall
current (turning back to usual units)
\begin{equation}
<J_i>=\frac{1}{2n+1}\frac{e^2}{h} \epsilon_{ij} E_j
\label{Jhall2}
\end{equation}
In a real sample, where impurities are present, 
$\vec{\nabla}\cdot\vec{E}\ne 0$. 
In this case we can decompose the electric field $\vec{E}$ in two
parts
\begin{equation}
\vec{E}=\vec{E'}+\vec{E}^{imp}
\end{equation}
where $\vec{\nabla}\cdot\vec{E}'=0$ and 
$\vec{\nabla}\cdot\vec{E}^{imp}=\rho_e$, with 
$\rho_e$  the external charge due to impurities. 
With these definitions we can rewrite (\ref{Jhall1}) in the 
following form: 
\begin{equation}
<J_i>=-i\frac{\pi}{\theta} \epsilon_{ij}E'_{j}
-i\frac{\pi}{\theta} 
\epsilon_{ij}\left( E^{imp}_j-\partial_j\int d^2y~G(x-y)
\vec{\nabla}\cdot\vec{E}^{imp}\right) 
\label{Jhall3}
\end{equation} 
It is simple to show that 
\begin{equation}
E^{imp}_j-\partial_j\int d^2y~G(x-y)
\vec{\nabla}\cdot\vec{E}^{imp}=0 .
\end{equation}
The second term of the l.h.s of the previous equation can be
interpreted as the electric bidimensional field created by a density
charge 
$\rho_e=\vec{\nabla}\cdot\vec{E}^{imp}$. 

Thus, from (\ref{Jhall3}) we have (in usual units)
\begin{equation}
<J_i>=\frac{1}{2n+1}\frac{e^2}{h}\epsilon_{ij} E'_j
\label{Jhall4}
\end{equation}
So, the quantization of the Hall conductivity for $\nu=\pi/\theta$ 
is {\em exact} in this
model  and {\em does not depend on any impurity distribution}. In other
words, the transverse current only ``sees'' divergenceless fields.
System dynamics and random impurities only affect the edge states of
the system, and this the reason for the exact quantization of the Hall
conductance. This result generalize Prange's result \cite{prange} for
a $\delta-\mbox{type}$ impurity, to the case of a general impurity
distribution in the context of the CSLG theory for the FQHE. 

\section{Dynamics of Density Fluctuations \label{DF}}

In the last section we calculated the electromagnetic current
and, from the resulting expression, we were able to understand the
role of the Chern-Simons term, of the edge states --- that are
responsible for the exact quantization of the conductivity --- and the
role of impurities, too. However, for a complete understanding of the
electromagnetic response, it is necessary to evaluate the polarization
tensor 
\begin{equation}
\Pi_{\mu\nu}(x,y)=\frac{\delta^2\ln Z(A)}{\delta A^\mu(y)\delta A^\nu(x)}
\end{equation} 
The calculation of the functional derivatives is straightforward 
and the result is formally
expressed in terms of mean values of the density fluctuation
$<\delta\rho(x)\delta\rho(y)\ldots>$. In particular, the expression
for $\Pi^{00}$ is very simple
\begin{equation}
\Pi^{00}(x,y)=<\delta\rho(x)\delta\rho(y)>
\label{pi}
\end{equation}
were $\delta\rho(x)=\rho(x)-<\rho(x)>$

In order to explicitly calculate this type of objects we need to
evaluate the effective action for the density fluctuation
$\delta\rho$. The action (\ref{22}) is the appropriate one for the
study of the dynamics of density fluctuations around the ground state
of the model. The structure of the ground state depends on the
specific configuration of the external electromagnetic field. We will
specialize our present analysis to the case of a static and
homogeneous field, the relevant case to the study of the QHE. The
ground state for $\nu=\pi/\theta$ is homogeneous, and is given by
\begin{equation}
<\rho(x)>=\bar\rho=\frac{\pi}{\theta}B 
\end{equation} 
so that
$\rho(x)=\bar\rho+\delta\rho(x)$. The aim of the present section is to
obtain from (\ref{22}) an effective action for $\delta\rho(x)$ and to
extract from it the Feynman's rules to calculate mean values. We need
to perform a Taylor expansion of the action around the configuration
$\rho(x)=\bar\rho$. This expansion is straightforward, except for two
terms originated from integrating over the longitudinal component of
the gauge field. These terms are the kinetic term
$\partial_0\rho\left\{ \frac{1}{\vec{\nabla}\cdot(\rho\vec{\nabla})}
\right\} \partial_0\rho$, and the determinant obtained when
integrating over $\varphi$, namely,
$Tr\ln(\vec{\nabla}\cdot\frac{\rho}{m}\vec{\nabla})$. Let us analyze
these two terms carefully. We can write the kinetic term of
(\ref{22}) in the following way: 
\begin{equation}
I_K=\int d^3z_1d^3z_2~\delta\dot\rho(z_1){\cal K}(z_1,z_2)
\delta\dot\rho(z_2)
\label{Ik}
\end{equation} 
where
\begin{equation}
\vec{\nabla}_{z_1}\cdot
\left(\rho(z_1)\vec{\nabla}_{z_1}{\cal K}(z_1,z_2)\right)
=\delta(z_1-z_2)
\label{K}
\end{equation}        
($\delta\dot\rho$ means the derivative of $\delta\rho$ with respect to
time).
We have also used that the ground state density is static. 
Note that ${\cal K}$ depends on $\rho$ through the implicit equation 
(\ref{K}). Thus, we can expand the kernel ${\cal K}$ in powers of 
$\delta\rho$ in the following form, 
\begin{equation}
{\cal K}(z_1,z_2)=\left. {\cal K}(z_1,z_2)\right|_{\rho=\bar\rho}+
\int d^3x~\left. \frac{\delta {\cal K}(z_1,z_2)}{\delta \rho(x)}
\right|_{\rho=\bar\rho} \delta\rho(x)+\ldots
\label{taylor}
\end{equation} 
From  (\ref{K}) we see that 
\begin{equation}
\left. {\cal K}(z_1,z_2)\right|_{\rho=\bar\rho}=
\frac{1}{\bar\rho}G(z_1,z_2)
\label{constant}
\end{equation}
where $G(z_1,z_2)$ is the Green's function of the Laplacian 
(see eq. (\ref{12})). Also, we can show (see Appendix) that 
\begin{equation}
\left. \frac{\delta {\cal K}(z_1,z_2)}{\delta \rho(x)}
\right|_{\rho=\bar\rho}=
\frac{1}{\bar\rho^2}\vec\nabla G(z_1,x)\cdot\vec\nabla G(x,z_2)
\label{linear}
\end{equation}
Introducing (\ref{constant}) and (\ref{linear}) into (\ref{Ik}) we find
\begin{eqnarray}
I_K&=&\frac{1}{\bar\rho}\int d^3z_1d^3z_2~\delta\dot\rho(z_1)G(z_1,z_2)
\delta\dot\rho(z_2) \nonumber \\
&+&\frac{1}{\bar\rho^2}\int d^3z_1d^3z_2d^3x~
\vec\nabla G(z_1,x)\cdot\vec\nabla G(x,z_2)
\delta\dot\rho(z_1)\delta\dot\rho(z_2)\delta\rho(x)\nonumber\\
&+&\ldots 
\label{Ikf}
\end{eqnarray} 

The next interesting term to be considered is 

\begin{equation}
I_{det}=Tr\ln(\vec{\nabla}\cdot\frac{\rho}{m}\vec{\nabla}),
\label{det}
\end{equation}
which can be rewritten as
\begin{equation}
I_{det}=Tr\ln(\frac{\bar\rho}{m}\nabla^2+\vec{\nabla}\frac{\delta\rho}{m}
\vec{\nabla})
\end{equation}
Observe that this term is due to the interaction of the density field
with the longitudinal component of the gauge field. This process is
governed by low-momentum transfer, so the logarithmic term with two
derivatives can be neglected, and we obtain
\begin{equation}
I_{det} \approx Tr\ln(\frac{\bar\rho}{m}\nabla^2)+
Tr\ln(1+\frac{\delta\rho}{\bar\rho})
\end{equation}
The first term in this expression is a constant and can be absorbed
into the global renormalization factor of the partition function. The
other term can be expanded in powers of $\delta\rho$, yielding

\begin{equation}
I_{det}\approx -\frac{1}{2}\int d^3x~\frac{\delta\rho^2}{\bar\rho^2}+
\frac{1}{3}\int d^3x~\frac{\delta\rho^3}{\bar\rho^3}
\end{equation}

The linear term can be absorbed by a suitable renormalization of the
chemical potential; the only effect of the quadratic term is
redefining the two-body potential $V(x,y)$ in the low-momentum limit.
In the case of local potentials, this term may be used to redefine the
coupling constant, but it is irrelevant in the case of long-distance
interactions.

Collecting all the terms, we obtain the effective action for the
density fluctuations, 
\begin{eqnarray}
S&=&\frac{m}{2}\frac{1}{\bar{\rho}}
	\int d^3x d^3y~\delta\dot{\rho}(x) G(x-y)
\delta\dot{\rho}(y) \nonumber \\
 &+&\frac{1}{2}\int d^3x d^3y~\delta\rho(x) 
	\left\{\frac{1}{4m\bar{\rho}}\nabla^2\delta(x-y)-
	\left( V(x-y)-\frac{\bar{\rho}}{m}(\frac{\theta}{\pi})^2
G(x-y) \right) \right\} \delta\rho(y)
	\nonumber \\
 &+& \frac{1}{8m\bar{\rho}^2}\int d^3x~
	\vec{\nabla}\delta\rho(x)\cdot\vec{\nabla}
\delta\rho(x)\delta\rho(x)-
\frac{1}{3\bar\rho^{3/2}}\int d^3x~\delta\rho^3+\ldots
	\nonumber \\
 &-&\frac{1}{2m}\frac{\theta^2}{\pi^2}\int d^3x d^3y d^3z~
	\vec{\nabla}G(x-y)\cdot\vec{\nabla}
G(x-z)~\delta\rho(x)\delta\rho(y)\delta\rho(z)
	\nonumber \\
 &+& \frac{m}{2}\frac{1}{\bar{\rho}^2}
	\int d^3x d^3y d^3z~
	\vec{\nabla}G(x-z)\cdot\vec{\nabla}
G(z-y)~\delta\dot{\rho}(x)\delta\dot{\rho}(y)
	\delta\rho(z)
	\nonumber \\
 &-&\int d^2x d^2y~ \delta\rho(x) G(x-y)\vec{\nabla}\cdot\vec{E}
\label{ac}
\end{eqnarray}

As the fluctuation $\delta\rho$ is small, the quadratic term
of the effective action may be regarded as defining the dynamics of
the fluctuations, whereas the cubic terms standing for perturbations
to such dynamics. In this sense, the propagator of the density
fluctuation in Fourier space may be written as
\begin{equation}
<\delta\rho(\omega,{\bf k})\delta\rho(-\omega,{\bf -k})>=
\frac{\frac{2\bar{\rho}}{m}{\bf k}^2}
{\omega^2-\left\{ \left(\frac{{\bf k}^2}{2m} \right)^2 +
\frac{\bar{\rho}}{m}{\bf k}^2 V(-{\bf k}^2)+ 
\frac{\bar{\rho}^2}{m^2}(\frac{\theta}{\pi})^2\right\}}
\label{prop}
\end{equation}
This propagator has very interesting information encoded.
For example, the compressibility of the system is defined as: 
\begin{equation}
\kappa=\lim_{{\bf k}\rightarrow 0} \Pi_{00}(0,{\bf k})~~~~~.
\label{comp}
\end{equation}
From (\ref{pi}) and the propagator (\ref{comp}) we have for all the
potentials with the general form $V(-{\bf k}^2)\propto {\bf
k}^{\alpha}$, 
\begin{equation}
\kappa=\lim_{{\bf k}\rightarrow 0} -\frac{\frac{2\bar{\rho}}{m}{\bf k}^2}
{ \left(\frac{{\bf k}^2}{2m} \right)^2 +
\frac{\bar{\rho}}{m}{\bf k}^2 V(-{\bf k}^2)+ 
\frac{\bar{\rho}^2}{m^2}(\frac{\theta}{\pi})^2}=0
\label{incomp}
\end{equation}
showing that the system is in an incompressible state.

The relation of this incompressibility to boson superconductivity
was first shown in ref.\ \cite{zang}. In the context of the present
work, incompressibility depends on two facts. The first one is the
existence of a gap in the spectrum of excitations. However, the gap
existence is a
necessary but not sufficient condition
to determine the incompressibility of the
ground state. The second fact is the presence of a ${\bf k}^2$ factor
in the numerator of the propagator. The gap in the spectrum is opened
by the transverse gauge field fluctuations and its value
$\delta\Delta=\frac{\bar{\rho}^2}{m^2}(\frac{\theta}{\pi})^2$
coincides with those found in refs. \cite{zang}, \cite{fradking} and
\cite{boyan}. This gapfull excitation corresponds to the cyclotron
mode excitation, and was identified as an inter-Landau level
excitation. The ${\bf k}^2$ factor in the numerator is induced by the
longitudinal gauge field fluctuations that affect the dynamics of the
density fluctuations. Remember that these fluctuations are completely
equivalent to the phase fluctuations of the matter field. 

It is useful to write the propagator (\ref{prop}) in the following
form:
\begin{equation} 
<\delta\rho(\omega,{\bf k})\delta\rho(-\omega,{\bf -k})>=
\frac{\pi}{\theta}{\bf k}^2 \left( \frac{1}{\omega-\omega_k}-
\frac{1}{\omega+\omega_k} \right)+ O({\bf k}^3)
\label{kohon} 
\end{equation}
with 
\begin{equation} 
\omega_k=\sqrt{ \left( \frac{{\bf k}^2}{2m}
\right)^2+ \frac{\bar\rho}{m} {\bf k}^2V(-{\bf k}^2)+
\frac{\bar{\rho}^2}{m^2} (\frac{\theta}{\pi}) }.
\label{dispertion}
\end{equation} 
This shows that our approach is consistent with Kohon's theorem
\cite{kohon}. This theorem asserts that in a planar
translation-invariant system submitted to a perpendicular magnetic
field. The density-density correlation function up to order ${\bf
k}^2$
must have a gap equal to $\omega_c=\frac{\bar\rho B}{m}$. This is an
exact result, independent of the microscopic details of the system.
Thus, a consistent calculus of radiative corrections must take into
account that constraint at each order of the perturbation expansion,
in addition to the vertices read from (\ref{ac}). Note that this
result is valid only for two-body potentials that vanish as
$r\rightarrow\infty$. For example, a logaritmic potential $V(-{\bf
k}^2)\propto 1/{\bf k}^2$ modifies the gap, as can be confirmed from
(\ref{dispertion}).

The dispersion relation (\ref{dispertion}) coincides with that
calculated in \cite{zang} and \cite{fradking} in the low-momentum
limit. For the Coulomb potential $V(-{\bf k}^2)\propto 1/|{\bf k}|$
the dispersion is linear, as it was first suggested in
ref.\cite{halpering}. Moreover, equation (\ref{dispertion}) coincides
with the dispersion relation of quasiparticles in an anyon
superfluid\cite{boyan}. It is interesting to note that, in ref.\
\cite{boyan} the model is written in terms of gauge invariant
operators 
\begin{equation} 
\Phi(x)\equiv \phi(x) e^{i\varphi(x)}~~~~~~~
\Phi(x)^*\equiv e^{-i\varphi(x)}\phi(x) 
\end{equation} 
where $\varphi(x)$ is the
longitudinal component of the gauge field given by (\ref{decomp}). The
theory is canonically quantized and then, the excitations of
quasiparticles are calculated in the Bogoliubov approximation. In our
functional approach, the gauge invariant degree of freedom is chosen
to be the particle density $\rho(x)$ and integration over all the
remaining fields is carried out. Studying density fluctuations, we
find that the dispersion relation of the cyclotron excitation is
exactly the same as the former. This should mean that the term $({\bf
k}^2/2m)^2$ of (\ref{dispertion}) is closely related to fluctuations
of gauge invariant objects. 

It is well known that this model presents another type of 
excitations called {\em topological vortices}. The typical asymptotic 
behavior of a vortex is (for $|x|\rightarrow \infty$)
\begin{eqnarray}
\phi(x)&=&\bar{\rho}^{1/2}e^{i\gamma(x)} \\
a_i(x)&=&\partial_i\gamma(x)=\epsilon_{ij}\frac{x_j}{|x|^2}
\end{eqnarray} 
where $\gamma$ is the azimutal angle. 
Finite energy requirements conduce to flux quantization 
\begin{equation}
\oint \vec{a}\cdot d\vec{s}=2\pi
\end{equation}
and due to the Chern-Simons constraint the charge of the vortex is
also quantized. Explicit solutions to the classical equations of
motion were found in several references
\cite{boyan}\cite{zang2}\cite{ezawa}. Here, we would like to find an
effective action for the particle density containing the vortex
dressed with quantum fluctuations of the gauge field. The effective
action (\ref{ac}) has lost its topological information and contains no
vortex. So far, we worked out all the functional integrations by
implicitly fixing trivial boundary conditions on the fields. In order
to obtain a density representation for the effective action in the
presence of a vortex, though, we must fix non-trivial boundary
conditions when performing the functional integrals.
 
In (\ref{GFT}) we fixed the gauge by imposing
$\delta\left(\theta(x)\right)$, and we have shown that this is
equivalent to choosing Coulomb's gauge. However, this is not the only
possible choice. Supose we allow a singular gauge transformation, 

\begin{eqnarray}
\theta(x)&\longrightarrow &\theta(x)+\Lambda(x)+\gamma(x) \label{gauge3}\\
a_\mu(x)&\longrightarrow &  a_\mu(x)+\partial_\mu\Lambda(x) +
\partial_\mu\gamma(x) 
\label{gauge4}
\end{eqnarray} 
($\gamma$ is the azimutal angle)  
and choose $\Lambda=-\theta$. This transformation fixes the phase to
the value $\theta=\gamma$ in the action, and induces a longitudinal
gauge component $a^L_i=\partial_i\theta+\partial_i\gamma$. Thus, we
can change the integration variables from $\theta$ to $a^L_i$.
Retracing the steps that lead from (\ref{GFT}) to (\ref{fixing}), we
conclude that we can also choose the gauge fixing functional to be
\begin{equation}
G_F(\theta,a_i)=\delta(\theta-\gamma)
\end{equation}
This gauge fixing automatically establishes topological boundary
conditions for the phase or, equivalently, for $a_i^L$. A similar
technique was also used in \cite{higs}, in the context of the abelian
Higgs model. 

With these boundary conditions, the action in terms of density fluctuations
takes the form 

\begin{equation}
S_V(\rho)=S_{eff}(\rho)+S_\gamma(\rho)
\label{sv}
\end{equation}
where $S_{eff}$ is the effective action with trivial boundary 
conditions given by (\ref{22}) and
\begin{equation}
S_\gamma(\rho)=\frac{1}{m}\int d^3x~\left(    
-\frac{1}{2}\rho|\vec{\nabla}\gamma|^2+\rho\vec{a}\cdot\vec{\nabla}\gamma
\right). 
\end{equation}
Integrating over the gauge field, we obtain
\begin{equation}
S_\gamma(\rho)=-\frac{1}{4m}\int d^3x~    
\rho|\vec{\nabla}\gamma|^2 + \frac{1}{m}\frac{\pi}{\theta}\epsilon_{ij}
\int d^3xd^3y~ \partial_i\rho(x)G(x-y)\gamma(x)\partial_j\rho(y)
\label{sgamma}
\end{equation}
The first term of this equation comes from the integration of the
longitudinal fluctuation of the gauge field, and can be interpreted as
the interaction with an external charge density, as it can be
rewritten in the form 
\begin{equation}
\frac{1}{4m}\int d^3x~    
\rho|\vec{\nabla}\gamma|^2=\frac{1}{m}\int d^3xd^3y~ \rho(x)G(x-y)\rho_e(y)
\end{equation}
where $\rho_e(x)=\nabla^2|\vec{\nabla}\gamma|^2=1/|x|^4$.

The second term of (\ref{sgamma}) comes from the integration of the
transverse gauge field fluctuation; such fluctuation induces a new
two-body interaction relating density gradients in orthogonal
directions. Note that $S_\gamma$ is not translationally invariant
because $\gamma(x)$ is not well defined at the origin (the nucleus of
the vortex). Thus equation (\ref{sv}), together with
(\ref{22}) and (\ref{sgamma}), is the effective action for the density
charge of bosons coupled to Chern- Simons gauge field in the presence
of a vortex centered at the origin of coordinates. It would be
possible, upon minimizing this action, to obtain a vortex profile
dressed with the quantum fluctuation of the gauge fields.

\section{Discussions and Conclusions \label{dis}}

We have considered in this paper a system of non-relativistic bosons
coupled to a Chern-Simons gauge field and an arbitrary external
electromagnetic field. We have used path integral techniques to deduce
an effective action in terms of the matter field density only. We
found the coupling of the charge density with the electromagnetic
field very interesting. The magnetic field acts as a ``background
density'', as it only enters the action through terms of the form
$\left(\rho(x)-\frac{\pi}{\theta}B(x)\right)$. The electric field only
couples to the density through its divergence
$\vec{\nabla}\cdot\vec{E}$, which in its turn is proportional to the
{\em external charges} (impurities, for example). So, in the absence
of impurities, the external electric field {\em does not couple with
the matter field}. This peculiar fact is due to the Chern-Simons
structure of the dynamical gauge field and it is the main reason for
the exact quantization of the transverse conductivity. 

The coupling of the gauge field with bosonic matter had two main
consequences on the density of the system. The integration over the
{\em longitudinal} gauge field produced a non-local dynamical term for
$\rho(x)$ and also a non-local interaction given by
\mbox{$Tr\ln(\vec{\nabla}\cdot\frac{1}{m}\rho\vec{\nabla})$.} The
integration over the {\em transversal} gauge field induced a two and a
three-body interaction terms. The induced two-body interaction is the
basic characteristic of a fluctuation of the transversal degrees of
freedom of the Chern-Simons field, and it is responsible for opening a
gap in the spectrum of excitations. Moreover, the longitudinal
fluctuations are also non-trivial. In particular the non-local
dynamical term modifies the propagator of density fluctuations,
leading, in addition to the appearence of a gap, to an incompressible
ground state. Another interesting aspect of this action is that, apart
from the interaction terms, a pure Chern-Simons term in the external
electromagnetic field is factored out. This fact is important in the
quantization of the transversal conductivity. 

Using these properties of the action, we were able to build an exact
expression for the conserved $U(1)$ current of the model. In
particular, we have explicitly shown that, in the case of a
homogeneous and static magnetic field, the quantization of the
transverse conductivity is exact, {\em even in the presence of any
type of impurities}. Thus, we generalized Prange's result
\cite{prange} for the FQHE to any kind of impurity distribution,
in the context of the CSLG theory. This development is based on the
fact that all the microscopic dynamics of the model and the coupling
with impurities are entirely contained in an automatically conserved
current (topological current) that does not contribute to the total
current. In the case of an infinite system, this topological density
current is zero. In a finite system the edge states exactly cancel
out the
contribution of this density to the total current. Moreover, we have
shown that the transverse current only ``sees'' divergenceless field,
thus canceling any contribution from external charges. 
 
We have computed the propagator of density fluctuations, using our
effective action. The propagator shows a gap in the excitation
espectrum that coincides with the inter-Landau level excitations
calculated previously using semiclassiclal approximations \cite{zang}.
The dispersion relation of this excitation coincides with that
calculated in ref. \cite{boyan} by a gauge-invariant canonical
formalism in the framework of anyon superconductivity. This
observation suggests that the form of the dispersion relation is
related to fluctuations of  gauge invariant objects. Of course, in
the low-momentum (long-distance) limit, the dispersion relation is
linear (as noted in \cite{halpering}) and coincides with all the other
approaches in this limit. 
  
We have also analyzed the structure of the effective action in the
presence of a vortex excitation. The main difference in this case is
that the integration over the longitudinal gauge field produces an
interaction with an induced external density charge of the form
$\rho_e=1/|x|^4$. The integration over the transversal gauge field
induces a new two-body interaction between the gradients of the
density in orthogonal directions. Upon minimizing this action we
should obtain the density profile of the vortex dressed with the
quantum fluctuation of the gauge field. In order to actually minimize
this action, we must face the problem of solving a system of a
non-local and non-linear integro-differential equations. This work is
in progress and will be presented elsewhere. 

The action $S_V$ should not be confused with the dual actions
developed in references \cite{dual1} and \cite{dual2}. Those dual
actions represent vortex densities, whereas the action in the present
work represents the particle density in the presence of a vortex. In
order to obtain an action for vortex excitation with our formalism, we
would have to introduce singularities into configuration space in a
way similar to that of ref. \cite{skyrmion}. This study is under
development and it is certainly beyond the scope of this paper.

\section*{Acknowledgements}
The Conselho Nacional de Pesquisa e Desenvolvimento(CNP$q$---Brazil),
and SR2-UERJ are gratefully acknowledged for financial support.

D.\ G.\ B.\ would like to acknowledge Prof.\ L.\ E.\ Oxman for useful
discussions.

\appendix

\section{Appendix}
\subsection{Expansion of the Kinetic term} 

In order to deduce the dynamics of density fluctuations, we must
expand the kinetic term of (\ref{22}) by performing a functional Taylor
expansion around the function $\rho(x)=\bar\rho$. Here, we would like
to sketch the principal steps in  developing such expansion. As
indicated in (\ref{Ik}) and (\ref{K}), the kinetic term of (\ref{22})
can be written as
\begin{equation}
I_K=\int d^3z_1d^3z_2~\delta\dot\rho(z_1){\cal K}(z_1,z_2)
\delta\dot\rho(z_2)
\label{Ika}
\end{equation} 
where
\begin{equation}
\vec{\nabla}_{z_1}\cdot
\left(\rho(z_1)\vec{\nabla}_{z_1}{\cal K}(z_1,z_2)\right)
=\delta(z_1-z_2)
\label{Ka}
\end{equation}
The kernel ${\cal K}$ is a functional of $\rho(x)$ given impicitly by
eq. (\ref{Ka}). Putting $\rho(x)=\bar\rho+\delta\rho(x)$ and expanding
up to the first order in $\delta\rho$, we find 
\begin{equation}
{\cal K}(z_1,z_2)[\delta\rho]
=\left. {\cal K}(z_1,z_2)\right|_{\rho=\bar\rho}+
\int d^3x~\left. \frac{\delta {\cal K}(z_1,z_2)}{\delta \rho(x)}
\right|_{\rho=\bar\rho} \delta\rho(x)+\ldots
\label{taylora}
\end{equation} 
Calculation of the first term is straightforward. From (\ref{Ka}),
we have
\begin{equation}
\vec{\nabla}_{z_1}\cdot
\left(\bar\rho\vec{\nabla}_{z_1}
\left.{\cal K}(z_1,z_2)\right|_{\rho=\bar\rho}\right)
=\bar\rho\nabla^2_{z_1}\left.{\cal K}(z_1,z_2)\right|_{\rho=\bar\rho}
=\delta(z_1-z_2)
\label{1a}
\end{equation}
Thus, 
\begin{equation}
\left. {\cal K}(z_1,z_2)\right|_{\rho=\bar\rho}=
\frac{1}{\bar\rho}G(z_1,z_2)
\label{constanta}
\end{equation}
where $G(z_1,z_2)$ is the Green's function of the Laplacian operator. 

The second term of (\ref{taylora}) is more involved. Functionally
differentiating equation (\ref{Ka}) with respect to $\rho$ we find
\begin{eqnarray}
\lefteqn{
\frac{\delta~}{\delta\rho(x)}\vec{\nabla}_{z_1}\cdot
\left(\rho(z_1)\vec{\nabla}_{z_1}{\cal K}(z_1,z_2)\right)
=}\nonumber \\
&=&\vec{\nabla}_{z_1}\cdot
\left(\delta(z_1-x)\vec{\nabla}_{z_1}{\cal K}(z_1,z_2)\right)+
\vec{\nabla}_{z_1}\cdot
\left(\rho(z_1)\vec{\nabla}_{z_1}
\frac{\delta~}{\delta\rho(x)}{\cal K}(z_1,z_2)\right)=0
\label{ka1}
\end{eqnarray}
Observing that the operator 
$\vec{\nabla}_{z}\cdot\left(\rho(z)\vec{\nabla}_{z}\ldots\right)=
\int dy~{\cal K}^{-1}(z,y)\ldots$
we can rewrite equation (\ref{ka1}) as
\begin{equation}
\int dy~{\cal K}^{-1}(z_1,y)
\frac{\delta~}{\delta\rho(x)}{\cal K}(y,z_2)=
-\vec{\nabla}_{z_1}\cdot
\left(\delta(z_1-x)\vec{\nabla}_{z_1}{\cal K}(z_1,z_2)\right)
\label{ka2}
\end{equation}
Inverting this equation we find 
\begin{equation}
\frac{\delta~}{\delta\rho(x)}{\cal K}(z_1,z_2)=
-\int dy~{\cal K}(z_1,y)\left\{ \vec{\nabla}_{y}\delta(y-x)\cdot
\vec{\nabla}_y{\cal K}(y,z_2)+\delta(y-x)\nabla^2_y{\cal
K}(y,z_2)\right\}
\label{ka3}
\end{equation}
and integrating the $\delta-\mbox{functions}$ we finally obtain
\begin{equation}
\frac{\delta~}{\delta\rho(x)}{\cal K}(z_1,z_2)=
\vec{\nabla}_x{\cal K}(z_1,x)\cdot\vec{\nabla}_x{\cal K}(x,z_2)
\label{ka4}
\end{equation}
This is a functional differential equation for ${\cal K}$. Finding its
solutions for arbitrary $\rho(x)$ is a very difficult task, but for
our purposes only the special case $\rho=\bar\rho$ needs to be
considered. Using (\ref{constanta}) and (\ref{ka4}) we have
\begin{equation}
\left. \frac{\delta {\cal K}(z_1,z_2)}{\delta \rho(x)}
\right|_{\rho=\bar\rho}=
\frac{1}{\bar\rho^2}\vec\nabla G(z_1,x)\cdot\vec\nabla G(x,z_2)
\label{lineara}
\end{equation}
that, of course, coincides with equation (\ref{linear}).

\end{document}